\documentclass[]{spie}  %>>> use for US letter paper
\usepackage[]{graphicx}

\usepackage{booktabs}       % professional-quality tables
\usepackage{amsfonts}       % blackboard math symbols
\usepackage{siunitx}		    % \ang{}

\usepackage{amsmath}
\usepackage{subcaption}

\usepackage{algorithm}
\usepackage{algpseudocode}

\DeclareMathOperator*{\argmax}{arg\,max}
\DeclareMathOperator*{\argmin}{arg\,min}

\title{TextureWGAN: Texture Preserving WGAN with MLE Regularizer for Inverse Problems} 

\author{Masaki Ikuta and Jun Zhang
\skiplinehalf
Department of Electrical Engineering and Computer Science, \\ University of Wisconsin Milwaukee, Milwaukee, WI 53211, U.S.A.
}
 
\begin{document} 
  \maketitle 

%%%%%%%%%%%%%%%%%%%%%%%%%%%%%%%%%%%%%%%%%%%%%%%%%%%%%%%%%%%%%

\section{Introduction}

Inverse problems have become an active research area using deep learning, which has become popular since the inception of AlexNet in 2012. \cite{alex_net_paper, deep_learning_nature_paper} The most popular loss function in deep learning and inverse problems is the Mean Square Error (MSE). \cite{deep_learning_book} The MSE measures Euclidean distance between two images. The MSE loss is easy to calculate and it is differentiable, which is why it is popular in deep learning. \cite{deep_learning_book} It is also known that the MSE preserves pixel fidelity in inverse problems \cite{yang_paper, tomogan_paper} which is the most important requirement for medical imaging. While the MSE loss provides many benefits, it is also known that the resulting images are often too smooth. In medical imaging, clinicians express their concerns that de-noised images look unrealistic because de-noising algorithms make images too flat and diagnostic quality to assess images is degraded due to the different look and feel of post-processed images. \cite{kim_paper} The main cause of this problem is that the MSE loss measures the Euclidean distance for all pixels in each image and ignores the spatial information of the pixels such as image texture.

Changing the look and feel or image texture by an image processing algorithm is a major concern in medical imaging. \cite{kim_paper, eilaghi_paper, nailon_paper, funama_paper} Low contrast detectability in Computed Tomography is highly dependent upon image texture. \cite{funama_paper} The visibility of low contrast objects is impacted by the noise texture pattern more significantly than that of high contrast objects with the same size. \cite{funama_paper} In fact, the image texture of cancer is highly relevant to patient survival rates. \cite{funama_paper} Many clinicians use a blending factor to avoid image texture change in de-noising algorithms. \cite{kim_paper} This blended image has become popular because image texture is known to be preserved. But this blending method comes at the expense of degraded Peak Signal to Noise Ratio (PSNR) and Structure Similarity (SSIM).

In this paper, we propose a new method for inverse problems that uses a discriminator and a generator of a Wasserstein GAN (WGAN). \cite{wgan_paper, wgan_gp_paper} The WGAN minimizes the distance between the probability distribution of a true image data set and a generated image data set. The WGAN ensures that it can reduce noise and the underlying density distribution of the de-noised image is also close enough to the density distribution of the true image data set. The proposed method also uses a maximum likelihood estimation (MLE) regularizer to help the algorithm to keep pixel fidelity and make the resulting images look more genuine. Multiple loss functions are used in the regularizer. To utilize multiple loss functions in the regularization, an MLE approach is used to balance multiple losses in the proposed algorithm. This is because each loss function has a different unit scale. Also, some loss function is more reliable than others. An MLE regularizer is introduced to utilize multiple loss functions more effectively. Furthermore, we conduct an image texture analysis to see whether the proposed method preserves image texture. PSNR and SSIM analysis are also performed as many other inverse problem research projects have done.

The contribution of this paper is two-fold. We will show that this WGAN method can preserve image texture in inverse problems while PSNR and SSIM are kept at a high level. A convolutional neural network (CNN) with the MSE loss function will be used for comparison. Also, we will show the effectiveness as to how the MLE regularizer helps utilize multiple loss functions in the regularizer.

\section{Methods}

Inverse problems can be mathematically formulated by the following equation. \cite{adler_paper}

\begin{equation} \label{eq:equation2}
  \begin{aligned}
    y &= F(x) + \delta y,
  \end{aligned}
\end{equation}
where \(y \in Y\) is an observation and \(x \in X\) is the corresponding image to be reconstructed. \(X, Y\) are reconstructed space and measurement space that are generally Hilbert Spaces. \(F: X \rightarrow Y\) is a forward operator that maps X to Y. \(\delta y\) is a measurement noise. One way to solve inverse imaging problems that have attracted a lot of interest is to first find a rough initial estimate of $x$ by a simple/fast algorithm (e.g., pseudo inverse) and then improve on it with another/more sophisticated process. For example, for CT imaging, we could provide an initial reconstructed CT image by FBP, which is very fast but more prone to Poisson noise and other artifacts, and improve it by a deep neural network. In this work, we use this approach and use a WGAN to improve an initial FBP CT image where the FBP CT image is the input to the WGAN.

The optimized discriminator of WGAN \(D\) given any generator \(G\) is to maximize the following equation. \cite{wgan_paper, wgan_gp_paper}

\begin{equation} \label{eq:equation3}
  \begin{aligned}
    D_G^{*} &= \argmax_{D} \Big\{ \mathbb{E}_{X \sim p_r}[D(X)] + \mathbb{E}_{\hat{X} \sim p_g}[1-D(\hat{X})] \Big\} = \argmin_{D} \Big\{ \mathbb{E}_{X \sim p_g}[D(\hat{X})] - \mathbb{E}_{X \sim p_r}[D(X)] \Big\},
  \end{aligned}
\end{equation}
where \(p_r\) is the training data distribution (true data set) in a compact metric space and \(p_g\) is the generated distribution. \(X\) and \(\hat{X}\) are samples from \(p_r\) and \(p_g\) respectively.

WGAN uses Wasserstein distance, also known as Earth Mover's (EM) distance, to obtain the distance of equation (\ref{eq:equation3}). \cite{wgan_paper} In fact, the discriminator is called critic in WGAN because it cannot really discriminate between real and fake images in WGAN. This is one difference from GANs. WGAN also requires that the critic must lie within the space of 1-Lipschitz functions to keep the continuity and differentiability of its loss function where WGAN enforces through weight clipping. \cite{wgan_paper} However, weight clipping in WGAN leads to optimization difficulties. \cite{wgan_gp_paper} Gulrajani et al. \cite{wgan_gp_paper} proposed to use a gradient penalty to overcome this problem.

\begin{equation} \label{eq:equation4}
    D_G^{*} = \argmin_{D} \Big\{ \mathbb{E}_{\hat{X} \sim p_g}[D(\hat{X})] - \mathbb{E}_{X \sim p_r}[D(X)] + \mu \mathbb{E}_{\tilde{X} \sim p_i}[(\| \nabla_{\tilde{X}} D(\tilde{X}) \| - 1)^2] \Big\},
\end{equation}
where \(p_i\) is the sampling distribution that is sampled uniformly along straight lines between pairs of points sampled from the true data distribution \(p_r\) and the generated data distribution \(p_g\). \cite{wgan_gp_paper} Also, \(\mu\) is called the penalty coefficient and it was suggested to use \(\mu = 10\). \cite{wgan_gp_paper} This is how the critic is optimized.

Similarly, the optimized generator \(G\) of WGAN given any critic \(D\) is to minimize the following equation. \cite{wgan_paper, wgan_gp_paper}

\begin{equation} \label{eq:equation6}
  \begin{aligned}
    G_D^{*} &= \argmin_{G} \Big\{ \mathbb{E}_{X \sim p_r}[D(X)] + \mathbb{E}_{\hat{X} \sim p_g}[1 - D(\hat{X})] \Big\} = \argmin_{G} \Big\{ - \mathbb{E}_{\hat{X} \sim p_g}[D(\hat{X})] \Big\}.
  \end{aligned}
\end{equation}

To ensure pixel-level integrity and improve image look and feel, we use a regularizer where the MSE loss and a perception loss are employed, respectively. These two loss functions have been previously researched and they were found effective in inverse problems. \cite{yang_paper, tomogan_paper}

\begin{equation} \label{eq:equation7}
  \begin{aligned}
    G_D^{*} &= \argmin_{G} \Big\{ - \mathbb{E}_{\hat{X} \sim p_g}[D(\hat{X})] + \mathbb{R}(p_r, p_g) \Big\},
  \end{aligned}
\end{equation}

\begin{equation} \label{eq:equation8}
  \begin{aligned}
  \mathbb{R} (p_r, p_g) &= \lambda_1 \mathbb{E}_{\substack{\scriptscriptstyle X \sim p_r \\ \scriptscriptstyle \hat{X} \sim p_g}} [\| X-\hat{X} \|^2] + \lambda_2 \mathbb{E}_{\substack{\scriptscriptstyle X \sim p_r \\ \scriptscriptstyle \hat{X} \sim p_g}} [( \Psi_{\theta}(X) - \Psi_{\theta}(\hat{X}))^2],
  \end{aligned}
\end{equation}
where \(X\) and \(\hat{X}\) are pair images. \(\lambda_1\) and \(\lambda_2\) are regularization parameters. \(\Psi_{\theta}\) is a parametrized functional in a compact metric space. The first term is the MSE loss and the second term is the perception loss in this paper although this equation can be extended to other loss functions.

In this paper, we propose the Maximum Likelihood Estimation (MLE) regularizer to balance multiple loss functions. The detailed description for the MLE regularizer will be shown in the full paper in detail.

\section{Experiments}

We have tested our algorithm in super resolution, image de-noising and Computed Tomography reconstruction. Because of the limited space in this abstract, only CT reconstruction results are shown. In CT images, we used an LIDC/IDRI data set. \cite{lidc_idri_data_set} We used 1041 images for training and 98 soft tissue-rich images for evaluation. All images were down-scaled from 512x512 to 256x256 pixels for training efficiency. Each training step randomly chose images from the training data set. We used the ODL python package \cite{odl_software} to perform the forward projection and the backprojection for CT reconstruction. 

\begin{table}
  \begin{center}
    \caption{Quantitative evaluation of CT reconstruction (LIDC/IDRI)}
    \label{tab:luna_table1}
    \begin{tabular}{lSSSSS}
      \toprule % <-- Toprule here
      \textbf{Method} & \textbf{PSNR (db)} & \textbf{SSIM (\%)} & \textbf{rangefilt (\%)} & \textbf{stdfilt (\%)} & \textbf{entropyfilt (\%)}\\
      \midrule % <-- Midrule here
      Original	& N/A	& N/A	& 100.00	& 100.00	& 100.00			\\
      FBP			& 23.11	& 0.83	& 136.55	& 129.69	& 103.08		\\
      MSE100\%		& 28.42	& 0.94	& 74.34	& 71.36	& 119.17		\\
      MSE 50\%		& 26.33	& 0.90	& 98.10	& 92.78	& 121.93		\\
      NLM Filter		& 23.10	& 0.83	& 135.56	& 128.75	& 103.70		\\	
      TextureWGAN	& 28.06	& 0.93	& 83.69	& 80.08	& 121.49		\\
      \bottomrule % <-- Bottomrule here
    \end{tabular}
    \begin{tabular}{lSSSS}
      \toprule % <-- Toprule here
      \textbf{Method} & \textbf{Contrast (\%)} & \textbf{Correlation (\%)} & \textbf{Energy (\%)} & \textbf{Homogeneity (\%)}\\
      \midrule % <-- Midrule here
      Original		& 100.00	& 100.00	& 100.00	& 100.00	\\
      FBP			& 126.04	& 99.22	& 90.77	& 92.73	\\
      MSE 100\%		& 62.15	& 101.35	& 83.19	& 102.27	\\
      MSE 50\%		& 80.07	& 101.25	& 84.81	& 98.11	\\
      NLM Filter		& 125.42	& 99.25	& 90.97	& 92.87	\\
      TextureWGAN	& 73.05	& 99.98	& 87.06	& 100.43	\\
      \bottomrule % <-- Bottomrule here
    \end{tabular}
  \end{center}
\end{table}

We calculated PSNR, Structure Similarity (SSIM), first-order statistical texture analysis (rangefilt, stdfilt and entropyfilt) as well as second-order analysis (Contrast, Correlation, Energy and Homogeneity) to conduct the quantitative analysis shown in Table \ref{tab:luna_table1}. PSNR is used to measure pixel fidelity in generated images. SSIM is used to measure perceived differences in structure between two sets of images. The first-order and the second-order statistical texture analysis methods are used to measure texture differences. The first-order analysis is only conducted in pixel-level but the second-order analysis accounts for the spatial inter-dependencies among pixels. The details of the texture analysis methods are not described in this abstract due to the limited space and will be described in the full paper. All texture analyses were conducted by normalizing all resulting numbers by the results of the original noise-free images. This is why all numbers of texture analysis results of original noise-free images were 100\%. Other numbers were relative to the numbers of original noise-free images.

We evaluated TextureWGAN along with FBP, MSE 100\%, and MSE 50\% methods. MSE 100\% used UNet and the MSE as its loss function. MSE 50\% was created after averaging an FBP image and the corresponding MSE 100\% image. This blending image is popular because it often preserves image texture. We also used a non-local mean filter indicated as NLM Filter in the table. This filter is also known to preserve image texture.  

As we can see in the results in Table \ref{tab:luna_table1}, PSNR and SSIM of MSE 100\% and TextureWGAN are on the same level. The table also shows texture analysis results. Noisy images and MSE 100\% were degraded on both first-order and second-order texture results. On the other hand, TextureWGAN relatively maintains both first-order and second-order texture while keeping PSNR and SSIM high. These quantitative results show that TextureWGAN is capable of maintaining high pixel fidelity while preserving image texture.

\begin{figure}
  \centering
  \begin{subfigure}[b]{0.25\linewidth}
    \includegraphics[width=\linewidth]{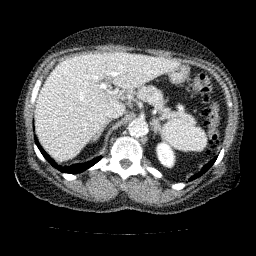}
    \caption{}
  \end{subfigure}
  \begin{subfigure}[b]{0.25\linewidth}
    \includegraphics[width=\linewidth]{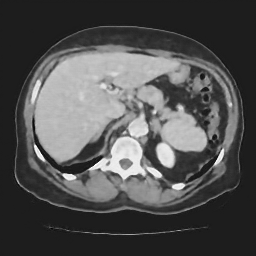}
    \caption{}
  \end{subfigure}
  \begin{subfigure}[b]{0.25\linewidth}
    \includegraphics[width=\linewidth]{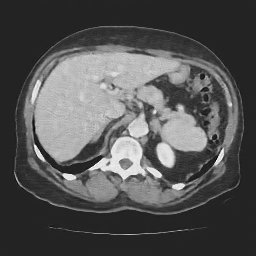}
    \caption{}
  \end{subfigure}
  
  \caption{CT reconstruction results on LIDC/IDRI data set. (a) filtered backprojection after Poisson and Gaussian noise added, (b) MSE 100\%, (c) TextureWGAN}
  \label{fig:luna_results}
\end{figure}

Fig.\ref{fig:luna_results} showed one of the results. We chose this image for validation since it has plenty of soft tissue pixels, which are generally more texture sensitive. As you can see, MSE 100\% over-smoothed the resulting image while TextureWGAN did not.

In summary, we have shown that the proposed method preserved image texture while it produced high PSNR and SSIM. Image texture analysis was conducted by visual inspection along with first-order and the second-order statistical texture analysis.

The proposed method is particularly suitable for medical imaging where both pixel fidelity and image texture need to be retained for clinical diagnostics. We plan to use more loss functions to try to maximize the performance of the proposed method.

\end{document}